\documentclass[twocolumn,superscriptaddress, showpacs] {revtex4-1}
\usepackage{graphicx}
\usepackage{amsmath}

\begin{document}

\preprint{AIP/123-QED}

\title[]{Long range universal quantum computation in large-size coupled cavity array independent of cavity number}

\author{Xiao-Qiang Shao}
\email{xqshao@yahoo.com.}
\affiliation{
School of Physics, Northeast Normal University,
Changchun 130024, People's Republic of China
}%
\affiliation{%
Centre for Quantum Technologies, National University of Singapore, 3 Science Drive 2, Singapore 117543
}%

\author{Tai-Yu Zheng}%
 \affiliation{
School of Physics, Northeast Normal University,
Changchun 130024, People's Republic of China
}%

\author{Jia-Bin You}
 \affiliation{%
Centre for Quantum Technologies, National University of Singapore, 3 Science Drive 2, Singapore 117543
}%
\author{C. H. Oh}
 \affiliation{%
Centre for Quantum Technologies, National University of Singapore, 3 Science Drive 2, Singapore 117543
}%
\author{Shou Zhang}
 \affiliation{%
Department of Physics, College of Science,
Yanbian University, Yanji, Jilin 133002, People's Republic of China
}%

\date{\today}

\begin{abstract}
We present a new approach for implementing a $\sqrt{\rm{swap}}$ gate between two spatially far apart sites connected by a large-size  coupled cavity array as quantum bus. The duration is only related to the parity of cavity number but independent of a specific number of cavity, thus it is possible to process quantum information in an arbitrary long distance in principle without time varied. Referring to the recent experimental progresses on coupled-cavity array, we also make an assessment of the scalability and take the cavity number $N=5$ as an example to illustrate the robustness of our proposal via quantum process tomography.
\end{abstract}

\maketitle

One main task in quantum information processing (QIP) is to design a physical quantum computer that utilizes  superposition and entanglement to outperform traditional computers by a far greater order of magnitude \cite{ekert,Grover,jones,Gulde}. In a quantum computer,  the principle of universal quantum computation guarantees any quantum circuit can be synthesized by a series of
single-qubit coherent rotations together with certain particular entangling
operations. Thus considerable theoretical and experimental efforts have been devoted to simulation of two-qubit quantum unitary operation \cite{zheng,Jaksch,Li,sch,zheng1,Isen,wermer} since the pioneer works accomplished in ion-trap system \cite{cirac} and cavity quantum electrodynamics system  \cite{barenco,sleator}.

The emergence of coupled-cavity models have attracted much attention for they provide the means to overcome the problem of individual addressability and meets the requirement of distributed quantum computation, i.e. performing state transfer, entanglement generation, or quantum gate operations between two distant qubits.
In general, an array of coupled cavity consists of some optical cavities that photons are permitted to hop between neighboring cavities. As each cavity doped with one or more atoms,  many interesting physical phenomena are simulated, e.g. the strongly interacting polaritons are experimentally observed in a photonic crystal or coupled toroidal micro-cavities \cite{hart}.  Anisotropic Heisenberg spin-1/2 chain and higher spin chain are realized in Refs.~\onlinecite{hart1,chen,Cho}.  The transition from Mott state to superfluid state can be achieved via modulating the detuning between the hopping photon and the doped two level system \cite{bose}. Recently, two theoretical proposals for one way quantum computation are also put forward  \cite{Lin,Song}.
\begin{figure}
\scalebox{0.42}{\includegraphics{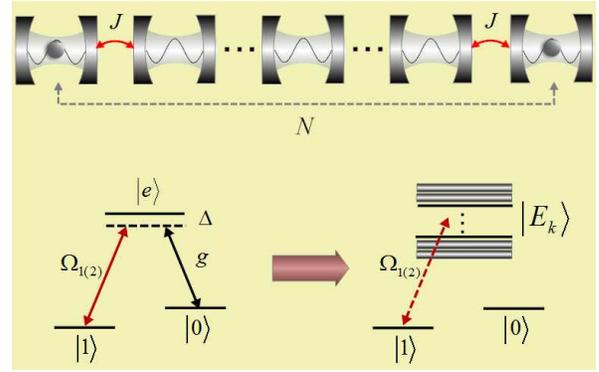}} \caption{\label{p00}
(Color online) Schematic of two three-level atoms doped in two end-sites of a coupled $N$-cavity arrays.
Each qubit is encoded into two lower-energy levels labelled as $|0\rangle$ and $|1\rangle$. The transitions between the levels
$|e\rangle\Leftrightarrow|0\rangle$ is coupled to the cavity
mode with the coupling constants $g$, and the
transitions $|e_{1(2)}\rangle\Leftrightarrow|1_{1(2)}\rangle$
 is driven by
a classical microwave pulses with the Rabi frequencies
$\Omega_{1(2)}$, $\Delta$ represents the
corresponding one-photon detuning parameter  and the photon can hop between two cavities with coupling strength $J$. $|E_k\rangle$ represents $k$th eigenenergy for the interaction between atoms and cavity array. }
\end{figure}

In this paper, we construct a large-size coupled-cavity model to perform a two-qubit $\sqrt{\rm{swap}}$ gate between two spatially far apart sites. This gate is a universal one and the operation time does not change as the increase of cavity number, i.e. we can carry out the quantum computation in an arbitrary distance in principle. This characteristic  may greatly reduce the complexity for experiment control.

The considered system consists of two $\Lambda$-type configuration qutrit
trapped in the end sites of a coupled-cavity array, as shown in Fig.~\ref{p00}.
Two low levels are used to encode qubit $|0\rangle$ and $|1\rangle$, respectively.
The transition from $|e\rangle$ to $|0\rangle$ is coupled to the cavity
mode with the coupling constant $g$, and the transition from $|e\rangle$ to $|1\rangle$ is driven by the classical fields with Rabi frequency $\Omega_{1(2)}$. The photon can hop between neighbor cavities with coupling strength $J$, and $\Delta$ represents the
corresponding one-photon detuning parameter. The Hamiltonian in the interaction picture reads ($\hbar = 1$):
$
H_I$=$\sum_{i=1,2}\Omega_i(|e_i\rangle\langle 1_i|+|1_i\rangle\langle e_i|)$
+$\Delta|e_i\rangle\langle e_i|$+$\sum_{j=1,2}g\big(a|e_j\rangle\langle
0_j|$+$|0_j\rangle\langle
e_j|a^{\dag}\big)$+$\sum_{k=1}^{N-1}J(a_k^{\dag}$
$a_{k+1}+a_ka^{\dag}_{k+1}).
$
For a two-qubit operation, the computation subspace is spanned by $\{|00\rangle, |01\rangle, |10\rangle, |11\rangle\}_a$, where the cavities are assumed initially in the vacuum state $|0,\dots,0\rangle_c$. Since the rotating wave approximation conserves the total number of excitations, we reclassify the above subspace according to different excited number, i.e. $|00\rangle_a|0,\dots,0\rangle_c$ decoupled to the dynamical evolution of system is
termed ``zero excitation" subspace; the  ``one-excitation" subspace includes states with either one  photon  in cavity or one atom in excited state $|e\rangle$; and the definition of ``two-excitation" subspace is in a similar way. Before proceeding, we briefly introduce the concept of quantum Zeno dynamics \cite{zeno} which enlightens us on this work. Suppose the dynamical evolution of a system is governed by
the Hamiltonian
$
{H}_K={H}+K{H}_c,
$
where ${H}$ is the Hamiltonian of the subsystem to be investigated,
and ${H}_c$ is an additional interaction Hamiltonian performing
the ``measurement", $K$ is the corresponding coupling strength.
In
a strong coupling limit $K\rightarrow\infty$, the system is dominated
by the effective Hamiltonian
$
 {H}_{eff}=\sum_nK\eta_n {P}_n+ {P}_n {H} {P}_n,
$
where $ {P}_n$ being the eigenprojection
of the $ {H}_c$ belonging to the eigenvalue $\eta_n$. In connection with the current model, the interaction between atoms and coupled-cavity array plays the role of continuous measurements on the interaction between atoms and the classical fields.
In what follows, we concentrate on the dynamical evolution of other three computation bases.

The states $|10\rangle_a|0,\dots,0\rangle_c$ and $|01\rangle_a|0,\dots,0\rangle_c$ belong to the ``one-excitation" subspace, after mapping the Hamiltonian of atom-cavity interaction to this subspace, we have a $(N+2)\times (N+2)$ matrix
\begin{eqnarray}\label{one}
H^{\rm I}_{single}=\left[\begin{array}{c c c c c c c c c}
\Delta_1 & g &0 & 0&\cdots &0 & 0 & 0 & 0 \\
g & 0 & J &  0&\cdots & 0 & 0 & 0 & 0 \\
0 & J & 0 &  J&\cdots & 0 & 0 & 0 & 0 \\
0 & 0 & J & 0 &\cdots & 0 & 0 & 0& 0  \\
\vdots & \vdots & \vdots &\vdots& \ddots&\vdots & \vdots & \vdots & \vdots \\
0 & 0 & 0 & 0 &\cdots & 0 & J & 0& 0  \\
0 & 0 & 0 & 0 &\cdots & J & 0 & J & 0  \\
0 & 0 & 0 & 0 &\cdots & 0 & J & 0& g  \\
0 & 0 & 0 & 0 &\cdots & 0 & 0 & g & \Delta_2  \\
\end{array}
\right]_{(N+2)\times (N+2)}.
\end{eqnarray}
For convenience we assume $g=J$, then Eq.~(\ref{one}) is equivalent to a tight-binding Hamiltonian with boundary impurities $\Delta_1$ and $\Delta_2$:
\begin{equation}\label{fer}
H_{\rm TBH}=\Delta_1 c_1^{\dag}c_1+\Delta_2 c_M^{\dag}c_M+J\sum_{i=1}^{M-1}(c_i^{\dag}c_{i+1}+c_ic_{i+1}^{\dag}),
\end{equation}
where $M=N+2$ throughout this paper unless otherwise specified.
Unlike the homogeneous spin model, the impurities in site 1 and $M$ break the translational symmetry, which leads to a complicated dynamical process. Fortunately, we may diagonalize Hamiltonian (\ref{fer}) via standard Green's function technique \cite{Green}.
Define $H_{\rm TBH}=H_{0M}+H_{1}$, where $H_{0M}=\Delta_2 c_M^{\dag}c_M+J\sum_{i=1}^{M-1}(c_i^{\dag}c_{i+1}+c_ic_{i+1}^{\dag})$ and $H_1=\Delta_1 c_1^{\dag}c_1$.
Then the Green's function of the total Hamiltonian $H_{\rm TBH}$ is
\begin{figure}
\scalebox{0.6}{\includegraphics{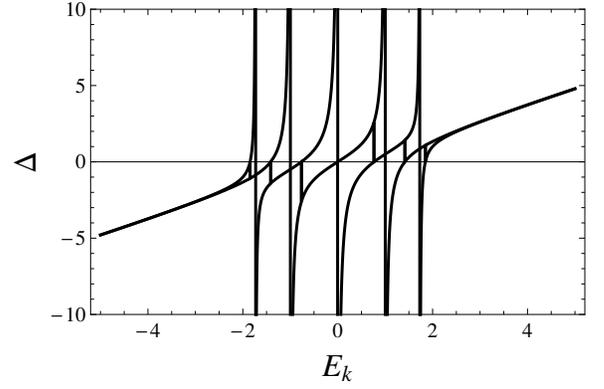}}\caption{\label{p2}
(Color online)  Plot of dispersion relation $E_k$ for 7-site tight binding Hamiltonian   with two equal boundary impurities $\Delta$ . All eigenenergies are bounded in the band $[-2J,2J]$ for zero defect. As the diagonal defect $|\Delta|$ increases, the two highest (lowest) eigenenergies move outside the band and become quasi-degenerate, and the other five eigenenergies approach the 5-site tight binding Hamiltonian without defects. All energies values are measured in units of $J=1$ for simplicity.}
\end{figure}
$
G=G_{0M}+G_{0M}|1\rangle\frac{\Delta_1}{1-\Delta_1 G_{0M}(1,1)}\langle 1|G_{0M},
$
where
$
G_{0M}=G_0+G_0\frac{\Delta_2}{1-\Delta_2 G_0(M,M)}G_0
$
corresponding to the Green's function of $H_{0M}$ and
\begin{equation*}\label{fred}
G_{0}=\frac{2}{N+1}\sum_{i=1}^{M}\sum_{j=1}^{M}\sum_{k=1}^{M}\frac{\sin[\frac{ik\pi}{M+1}]\sin[\frac{jk\pi}{M+1}]|i\rangle
\langle j|}
{E_k-2J\cos[\frac{k\pi}{M+1}]}
\end{equation*}
denotes the Green's function of Hamiltonian without impurity. Now we can extract all information about the eigenvalues and eigenfunctions of $H_{\rm TBH}$ from above Green's functions, e.g. the poles of $G$ disclose the spectrum of $H_{\rm TBH}$. In  Fig.~\ref{p2}, we plot the dispersion relation of a 7-site tight binding Hamiltonian ($M$=7) with two equal boundary impurities $\Delta$ via solving the algebraic equation $[1-\Delta G_{0M}(1,1)=0]$. Clearly, in the absence of impurity, all the eigenenergies are bounded in the band $[-2J,2J]$. The increase of positive $\Delta$ leads to
two highest eigenenergies move outside the band and become quasi-degenerate. As $\Delta$ increases further, other five eigenenergies asymptotically approach the values of a 5-site tight binding Hamiltonian without impurity, and the variation of a negative $\Delta$ plays a similar role. To sum up, the effect of  large impurities divide a $M$-site tight binding Hamiltonian into two subspaces, one includes two $\Delta$-dependent eigenstates composed by end sites and the other consists of the middle $(M-2)$ $\Delta$-independent eigenstates.

\begin{figure*}
\scalebox{0.26}{\includegraphics{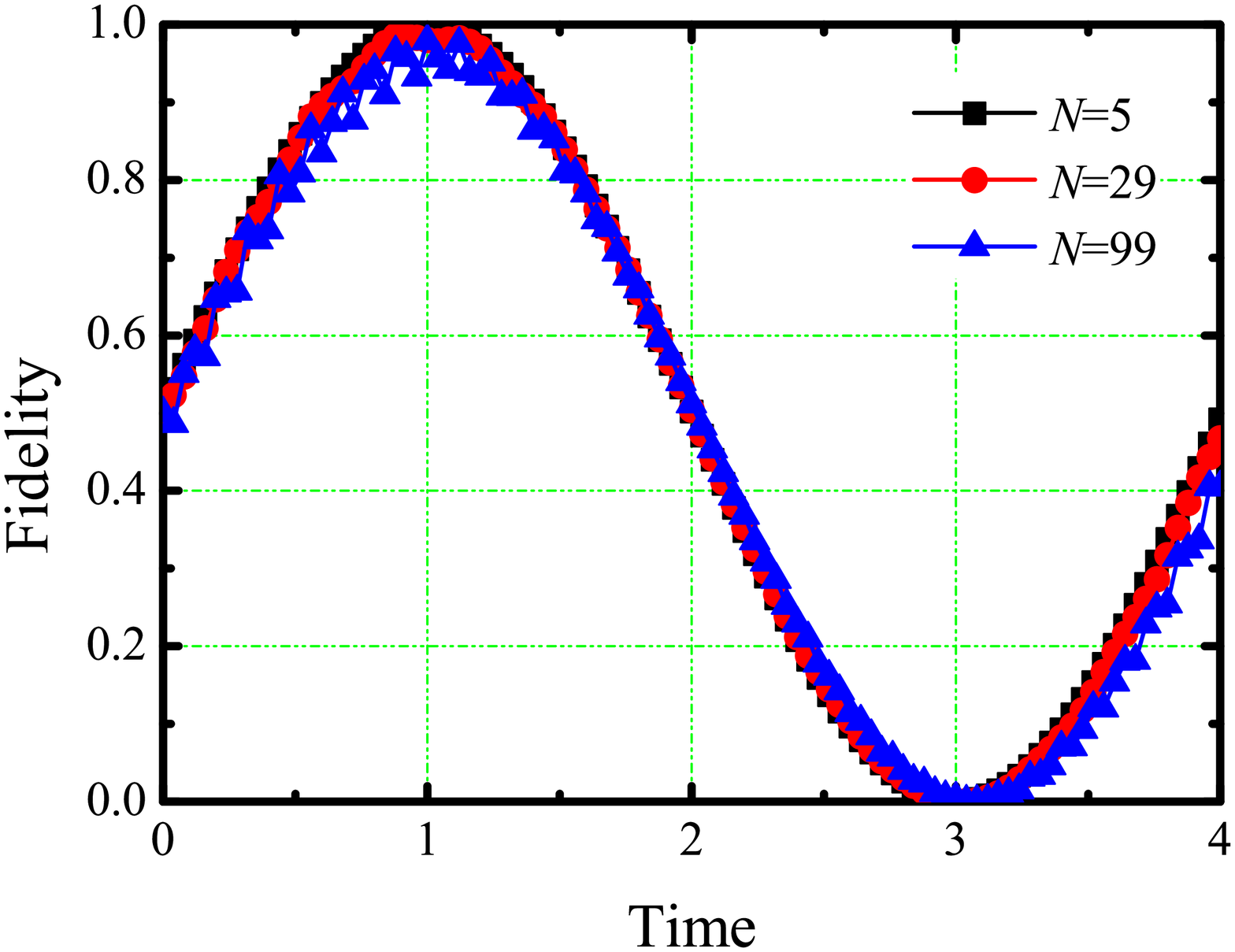}}
\scalebox{0.3}{\includegraphics{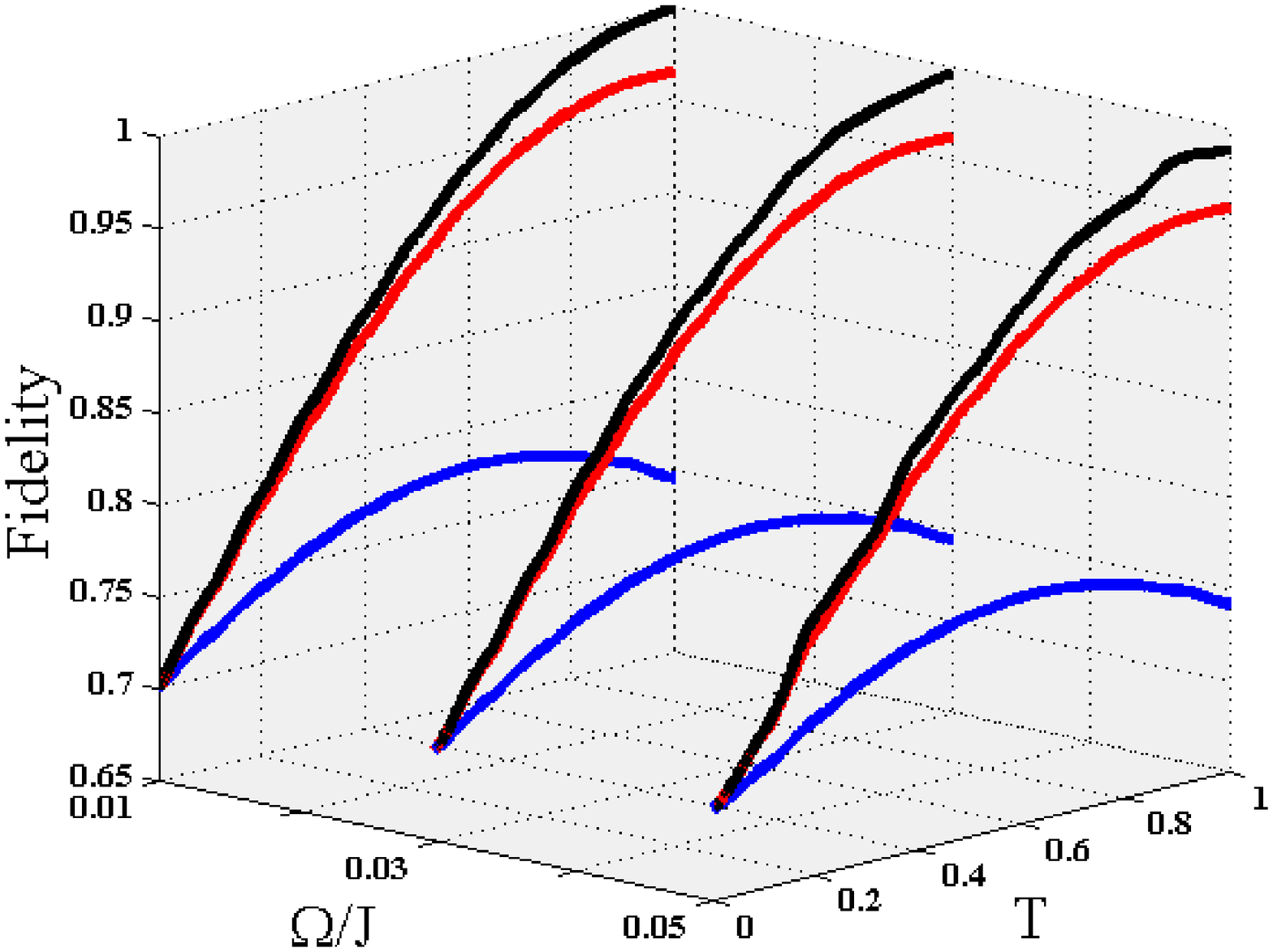}}\caption{\label{p3}
(Color online) (Left panel) Numerical simulating the fidelities of the entangling operation $|01\rangle\Rightarrow(1+i)/2|01\rangle+(1-i)/2|10\rangle$ for $N = 5$ (square), $N=29$ (circle), and $N=99$
(triangle). The time is reported in the unit of $T=\pi\Delta/(2\Omega^2)$. The maximal fidelity is obtained at $T$ for all cases, where the parameters are set as $\Delta=J=100/3\Omega$. (Right panel) Average fidelity $\overline{F}(t)$ vs. Rabi frequency $\Omega$ and the evolution time $T$ for 5-coupled-cavity system. In each group, the black curve represents an ideal case, while the red
and the blue lines corresponding decoherence parameters $\kappa=\gamma=0.01J$ and $\kappa=\gamma=0.1J$ respectively. }
\end{figure*}

With the full eigenenergies at hand, we are able to write down all the eigenfunctions with Lippman-Schwinger equation: $|E_k)=|E_k^0\rangle+(E_k-H_0)^{-1}|E_k)$, where $|E_k^0\rangle$ corresponds to the eigenstate for the unperturbed Hamiltonian $H_0$, and $|E_k)$ indicates the unnormalized eigenstate for the full Hamiltonian. In the language of Green's function, this equation can be reformulated as
$
|E_k)=[1-G_0(E_k)V]^{-1}|E^0_k\rangle,
$
where $V=\Delta_1 c_1^{\dag}c_1+\Delta_2 c_M^{\dag}c_M$ in our scheme and
$
|E^0_k\rangle$ =$\frac{\sqrt{2}}{\sqrt{M+1}}$$\{\sin[\frac{k\pi}{M+1}]$,$\sin[\frac{2k\pi}{M+1}],$$\cdots,$
$\sin[\frac{(M-1)k\pi}{M+1}],$$\sin[\frac{Mk\pi}{M+1}]\}$. Finally we obtain the
standard eigenfunctions of Eq.~(\ref{fer}) as
$
|E_k\rangle=\frac{|E_k)}{\sqrt{(E_k|E_k)}}=\{f_k^{1},f_k^{2},\cdots,f_k^{M-1},f_k^{M}\},
$
where $f_k^{i}=\langle i|E_k\rangle$ describes the probability amplitude of the site $|i\rangle$ in the eigenvector $|E_k\rangle$. Since the system
is mirror symmetric, the eigenvectors must be alternately
symmetric or antisymmetric, i.e. $\langle i|E_k\rangle=(-1)^{k+1}\langle \overline{i}|E_k\rangle$ for an ascending arrangement of $E_k$, where $\overline{i}$ means the mirror-conjugate site of $i$.

 Most interestingly, we find two relations exist in the current model which are independent of the specific ratio between $\Delta$ and $J$, i.e.
\begin{equation}\label{ww}
  \left\{
   \begin{aligned}
   &\sum_{k=1}^M\frac{f_k^{1*}\times f_k^M}{E_k}=(-1)^{\frac{M-1}{2}}\frac{1}{2\Delta} \\
   &\sum_{k=1}^M\frac{|f_k^{1}|^2}{E_k}=\sum_{k=1}^M\frac{|f_k^{M}|^2}{E_k}=\frac{1}{2\Delta} \\
   \end{aligned}
   \right.,
  \end{equation}
Eq~(\ref{ww}) is the crucial
results throughout this paper because it determines the interaction form of two end sites. Explicitly,
supposing the Rabi frequencies $|\Omega_1|$ and $|\Omega_2|$ are much weak compared with the values of $\{|E_k|, |E_k-E_{k^{'}}|\}$, we will always have an effective dipole-dipole interaction between $|10\rangle_a|0,\dots,0\rangle_c$ and $|01\rangle_a|0,\dots,0\rangle_c$ with a fixed strength $|\frac{\Omega_1\Omega_2}{2\Delta}|$ regardless of length of the chain, so is the case for the stark shifts of two states.

For the computation basis $|11\rangle_a|0,\dots,0\rangle_c$, it is a good approximation to expand
corresponding atom-cavity Hamiltonian into the single excitation subspace due to the weak excitation $|\Omega_{1(2)}|$, the stark shift of state $|11\rangle_a|0,\dots,0\rangle_c$ then can be canceled from the above two uncorrelated block matrices. Therefore the effective Hamiltonian governing the evolution of whole system reduces to
\begin{eqnarray}\label{fred}
H_{eff}&=&(-1)^{\frac{N+1}{2}}\frac{\Omega_1\Omega_2}{2\Delta}|10\rangle_a\langle01|+{\rm H.c.}\nonumber\\&&+\frac{\Omega_1^2}{2\Delta}|10\rangle_a\langle10|+\frac{\Omega_2^2}{2\Delta}|01\rangle_a\langle01|,
\end{eqnarray}
where we have discarded the term of cavities for they all stay in the vacuum state. Note Eq.~(\ref{fred}) is derived only under the assumption $|\Omega_{1(2)}|\ll\{|E_k|, |E_k-E_{k^{'}}|\}$, which has no restriction on the relation between $\Delta$ and $J(g)$. The dipole-dipole interaction between two end sites are caused by summation of $k$
independent virtual-photon-induced Raman transitions. To achieve the two-qubit $\sqrt{\rm swap}$ gate within the shortest time, the related parameters should satisfy $(-1)^{\frac{N-1}{2}}\Omega_1\Omega_2$=$\Omega_1^2$=$\Omega_2^2$, which signifies $(-1)^{\frac{N-1}{2}}\Omega_1$=$\Omega_2$=$\Omega$. In the left panel of Fig.~\ref{p3} we numerically simulate the fidelity of
state transforming $|01\rangle\Rightarrow(1+i)/2|01\rangle+(1-i)/2|10\rangle$ for $N = 5$, $N=29$, and $N=99$
with the full Hamiltonian, respectively. It shows under the given parameters $\Omega=0.03J=0.03\Delta$, the maximal fidelity can be achieved simultaneously at the very time $T=\pi\Delta/(2\Omega^2)$. Even for the long range  $N=99$, the fidelity remains above $97\%$. In principle, a much smaller $\Omega$ will result in a higher fidelity. Nevertheless, this may cost a much longer interaction time and render the system more susceptible to decoherence.

\begin{figure}
\scalebox{0.26}{\includegraphics{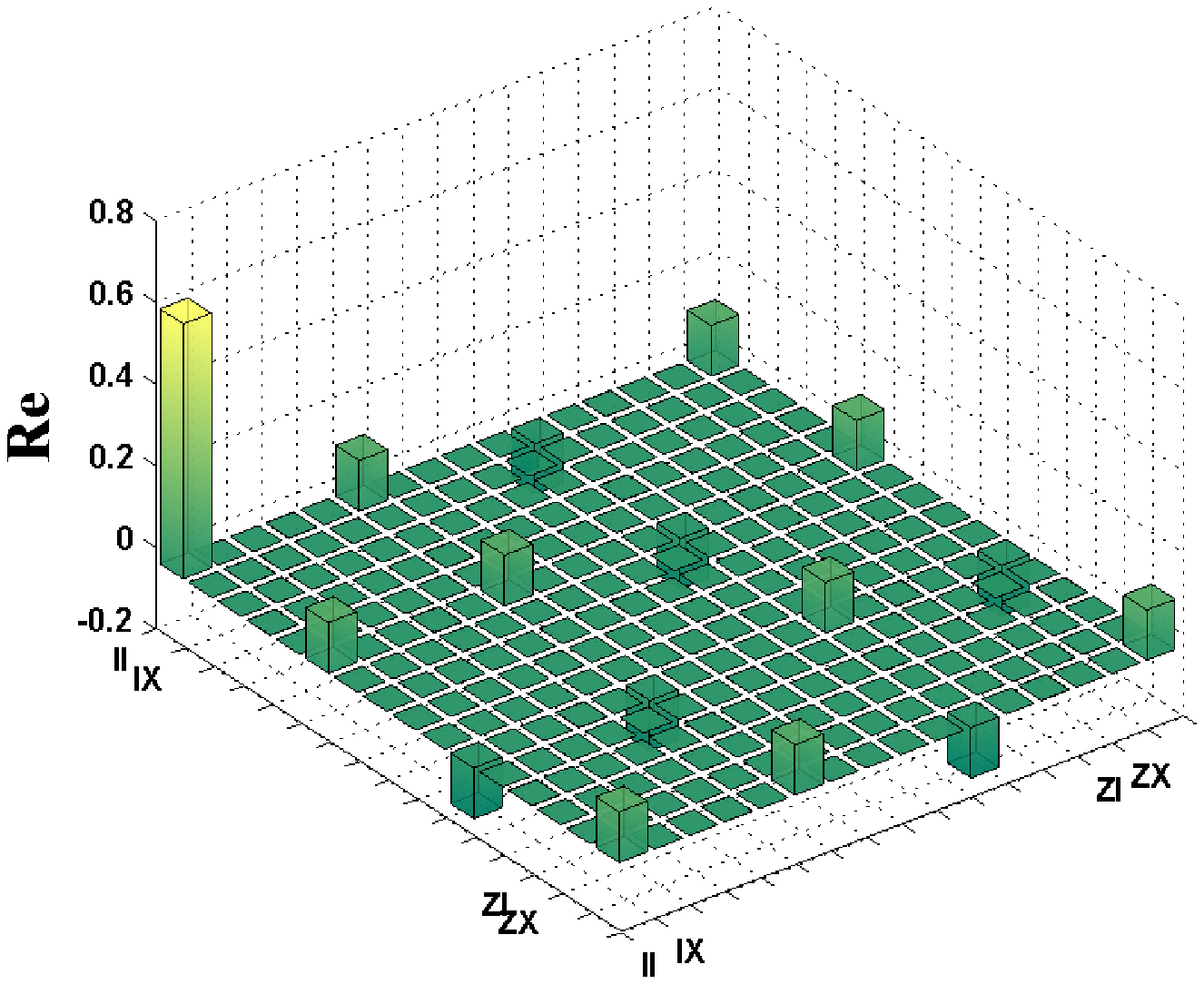}}
\scalebox{0.26}{\includegraphics{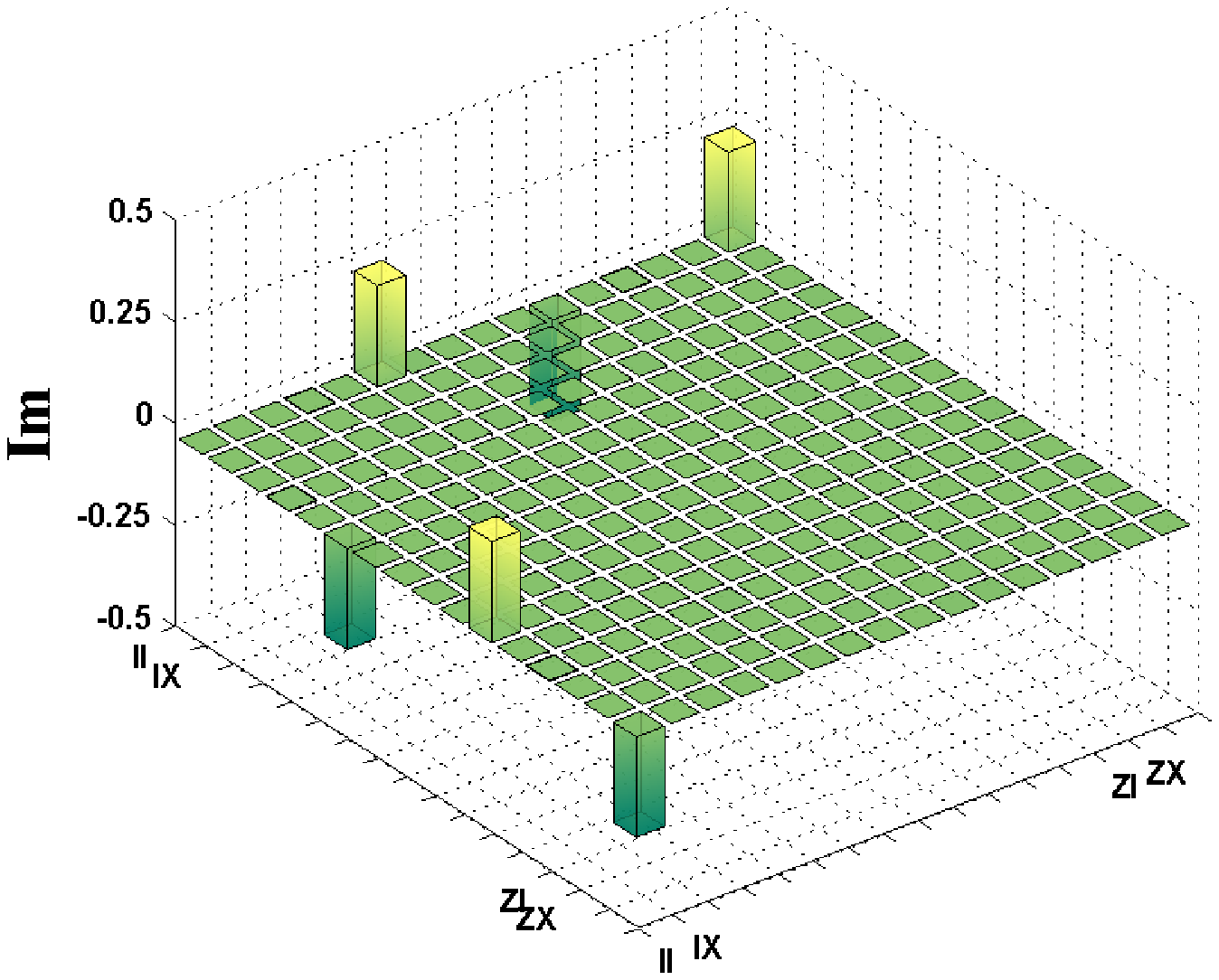}}\caption{\label{p4}
(Color online) The quantum process tomography of the $\sqrt{\rm swap}$ gate, which overlaps with an ideal $\chi$ matrix with 99.32\%, where the $chi$ matrix is measured in the modified Pauli basis $Y\rightarrow-iY$. }
\end{figure}

In the current model,  the master equation of the whole system
can be expressed by the Lindblad form \cite{scully}
$
\dot{\rho}=-i[H_{I},\rho]-\sum_{i=1}^{N}\frac{\kappa}{2}(a_i^\dag a_i\rho-2a_i\rho
a_i^\dag+\rho a_i^\dag
a_i)-\sum_{j=0,1}\sum_{k=1}^2\frac{\gamma_k^{ej}}{2}(\sigma_{ee}^k\rho-2\sigma_{je}^k\rho\sigma_{ej}^k+\rho\sigma_{ee}^k),
$
where $\kappa$ denotes the decay rate of cavity, $\gamma_k^{ej}$
represents the branching ration of the atomic decay from level
$|e\rangle_k$ to $|j\rangle_k$ and we assume
$\gamma_n^{e0}=\gamma_n^{e1}=\gamma/2$ for
simplicity. The performance of two-qubit $\sqrt{\rm swap}$ gate is evaluated via the definition of average fidelity
\cite{Measure,Mea1}
\begin{eqnarray*}\label{He}
\overline{F}(\varepsilon,U_{\rm \sqrt{swap}})=\frac{\sum_j{\rm tr}\big[U_{\rm\sqrt{swap}}U_j^\dag
U_{\rm \sqrt{swap}}^\dag \varepsilon(U_j)\big]+d^2}{d^2(d+1)},
\end{eqnarray*}
where $d=4$ for two qubits and $U_j$ being the tensor of Pauli
matrices  $II, IX,IY,\cdots, ZZ$, $U_{\rm \sqrt{swap}}$ being the ideal ${\rm \sqrt{swap}}$
gate and $\varepsilon$ being the trace-preserving quantum
operation obtained through our scheme. In the right panel of Fig.~\ref{p3}, we illustrate the average fidelity $\overline{F}(t)$  for a 5-coupled-cavity system under decoherence with three kinds of strength: $\kappa=0$, $\kappa=0.01J$ and $\kappa=0.1J$.
In the ideal case, the fidelities are $99.97\%$, $99.69\%$ and $98.80\%$ corresponding $\Omega=0.01J$,  $\Omega=0.03J$ and  $\Omega=0.05J$, which agree well with our previous statement. Although the fidelity decreases as the increase of $\kappa$, a relatively high fidelity is still available in the range $\kappa<0.01J$. To completely characterize the dynamical process, we give the quantum process tomography of the $\sqrt{\rm swap}$ gate in Fig.~\ref{p4}  with the parameters ($g,\gamma,\kappa)\sim(2.5\times10^9,1.6\times10^7,4\times10^5$) Hz referred  to a recent experiment about large-scale arrays of ultrahigh-$Q$ coupled nanocavities \cite{Not}.
In the modified Pauli basis $Y\rightarrow-iY$, the overlap of $chi$ matrix between our scheme and the ideal one is $99.32\%$, which confirms the effectiveness of our assumption further. It should be pointed out that the current model is not limit to realization of $\sqrt{\rm swap}$ gate only. Eq.~(\ref{ww}) constructs the building block for cavity-number-independent long range quantum information processing, thus other forms of two-qubit gate such as CNOT gate and conditional $Z$ gate can also be implemented via modulating the classical fields acting on end atoms. For an even number of cavity, we find a similar relation as
\begin{equation}
  \left\{
   \begin{aligned}
   &\sum_{k=1}^M\frac{f_k^{1*}\times f_k^M}{E_k}=(-1)^{\frac{M}{2}}\frac{J}{\Delta^2-J^2} \\
   &\sum_{k=1}^M\frac{|f_k^{1}|^2}{E_k}=\sum_{k=1}^M\frac{|f_k^{M}|^2}{E_k}=\frac{\Delta}{\Delta^2-J^2} \\
   \end{aligned}
   \right.,
  \end{equation}
which means remote quantum computation can  be implemented in this case without time changed neither.

In summary, we have presented a scheme for long range universal quantum computation. This scheme utilizes a large-size coupled cavity array as a medium to induce the Raman coupling between two remote end sites, and the interaction time is independent of a specific number. The virtually excited photon process makes the scheme more robust against the typical decoherence parameters in cavity quantum electrodynamics.
We hope that our work may be useful for
the quantum information processing in the near future.

\begin{acknowledgments}
This work is supported by Fundamental Research Funds for the Central Universities under Grant Nos. 11QNJJ009 and 12SSXM001,  National
Natural Science Foundation of China under Grant Nos. 11204028 and 11175044, and National Research
Foundation and Ministry of Education, Singapore (Grant
No. WBS: R-710-000-008-271). X. Q. Shao was also supported in part by the Government of China through CSC.
\end{acknowledgments}

\end{document}